# LOCAL SOLUTION METHOD FOR NUMERICAL SOLVING OF THE WAVE PROPAGATION PROBLEM

V.E.Moiseenko

*National Science Center "Kharkov Institute of Physics and Technology", UA-310108 Kharkov, Ukraine; e-mail: moiseenk@ipp.kharkov.ua*

and

V.V.Pilipenko

*Scientific and Technological Center of Electrophysics, National Academy of Sciences of Ukraine, UA-310002, P.O.Box 8812, Kharkov, Ukraine*

## Abstract

A new method for numerical solving of boundary problem for ordinary differential equations with slowly varying coefficients which is aimed at better representation of solutions in the regions of their rapid oscillations or exponential increasing (decreasing) is proposed. It is based on approximation of the solution to find in the form of superposition of certain polynomial-exponential basic functions. The method is studied for the Helmholtz equation in comparison with the standard finite difference method. The numerical tests have shown the convergence of the method proposed. In comparison with the finite difference method the same accuracy is obtained on substantially rarer mesh. This advantage becomes more pronounced, if the solution varies very rapidly.

## 1. Introduction

In many fields of physics the problem of wave propagation and absorption in non-uniform media is under investigation. For example, in the controlled fusion problem the wave propagation phenomena are analyzed both analytically and numerically in application to the radio frequency plasma heating (e.g. [1]), the MHD plasma stability (e.g. [2]), the radio frequency plasma production (e.g. [3]). The problem of wave propagation is described, as a rule, by a set of differential equations. These equations are usually solved using the Fourier expansion or the discretization

(e.g. [4]). In the case of rapidly oscillating or exponentially increasing (decreasing) wave fields, employing of the standard discretization methods requires a very fine mesh. For example, to simulate the radio frequency field excitation in plasma of the LHD stellarator [3] the mesh with number of nodes $N$=20000 was used. This number will increase, if such modelling is performed in a larger device.

There exist a lot of problems when the coefficients of the set of differential equations which describe the wave propagation problem vary much slower than the solutions. In this case, in the regions of rapid oscillations of the wave field one can use the WKB solutions (e.g. [5]). However, using them, one cannot provide prescribed accuracy and, therefore, cannot control convergence of the calculations. For this reason the WKB approximation cannot be considered as a proper numerical method.

In the present paper we propose a new method for solving one-dimensional problem which improves the accuracy of numerical solution especially in the regions of rapid oscillations of the wave field. We call this approach as the local solution method. This method essentially exploits the fact that the solution of a system of linear differential equations can be represented as superposition of some linearly independent solutions of the homogeneous system and a specific solution. The main idea is to approximate this representation using certain basic functions in every mesh cell. There is wide freedom in choice of such functions. When using them in the form of polynomials we obtain results similar to the results of the standard mesh methods. Aiming at more exact representation of the solution in the regions of its rapid oscillation and exponential increase (decrease) we have chosen the basic functions in a polynomial-exponential form suggested by the form of the WKB solutions.

The method proposed is analyzed for the Helmholtz equation, the simplest equation describing wave processes. In section 2 we analyze theoretically three versions of the local solution method. In section 3 we present the results of numerical tests of the method in comparison with results obtained by the standard finite difference method.

## 2. Formulation of the local solution method

Consider the one-dimensional Helmholtz equation

$$\frac{d^2}{dx^2} y(x) + G(x) y(x) = R(x) , \qquad (1)$$

which is defined at the interval $x \in (x_l, x_r)$. We assume that the function $G(x)$ is not rapidly oscillating and has no breaks or singular points at this interval. To solve equation (1) numerically we introduce a mesh with the nodes $x_i$ where $i=1,2, \ldots ,n$. The strategy of the numerical solving of this equation is as follows. Within the segment $s_i$ ( $x_i \leq x \leq x_{i+1}$ ), the solution of equation (1) can be written in the form

$$y^{(i)}(x) = C_1^{(i)} y_1^{(i)}(x) + C_2^{(i)} y_2^{(i)}(x) + y_R^{(i)}(x) , \qquad (2)$$

where $y_1^{(i)}(x)$ and $y_2^{(i)}(x)$ are two linearly independent solutions of the homogeneous equation ($R(x)=0$) and $y_R^{(i)}(x)$ is a specific solution. Making use of the smallness of the segment considered, we assume that we can find approximations $\tilde{y}_j^{(i)}(x)$ (here $j=1, 2, R$) for these solutions with prescribed accuracy:

$$y_j^{(i)}(x) = \tilde{y}_j^{(i)}(x) + O(h^{m+1}) , \qquad (3)$$

where $h=x_{i+1}-x_i$, $m$ is the degree of the approximation. Since the approximate solutions $\tilde{y}_j^{(i)}(x)$ are known, in order to obtain the approximate solutions of equation (1), i.e. to determine the unknown coefficients $C_1^{(i)}$ and $C_2^{(i)}$,

$$\tilde{y}^{(i)}(x) = C_1^{(i)} \tilde{y}_1^{(i)}(x) + C_2^{(i)} \tilde{y}_2^{(i)}(x) + \tilde{y}_R^{(i)}(x) , \qquad (4)$$

at every segment we have to match the solutions and their derivatives at the internal mesh nodes:

$$\tilde{y}^{(i)}\Big|_{x=x_{i+1}} = \tilde{y}^{(i+1)}\Big|_{x=x_{i+1}} , \quad \frac{d\tilde{y}^{(i)}}{dx}\Big|_{x=x_{i+1}} = \frac{d\tilde{y}^{(i+1)}}{dx}\Big|_{x=x_{i+1}} . \qquad (5)$$

In this way we obtain $2(n-2)$ linear algebraic equations for $2(n-1)$ unknowns. Two equations still needed are to be obtained from two boundary conditions at the end point of the interval $x_l$ , $x_r$ . The resulting matrix of the described equation set has narrow band and, therefore, can be easily reversed. For the case of three segments and the boundary conditions posed at the opposite ends of the interval, the portrait of matrix is shown in equation 1.

|   |   |   |   |   |   |   |   |
|---|---|---|---|---|---|---|---|
| * | * | 0 | 0 | 0 | 0 | 0 | 0 |
| * | * | * | * | 0 | 0 | 0 | 0 |
| * | * | * | * | 0 | 0 | 0 | 0 |
| 0 | 0 | * | * | * | * | 0 | 0 |
| 0 | 0 | * | * | * | * | 0 | 0 |
| 0 | 0 | 0 | 0 | * | * | * | * |
| 0 | 0 | 0 | 0 | * | * | * | * |
| 0 | 0 | 0 | 0 | 0 | 0 | * | * |

*Figure 1.* The portrait of the matrix of the system.

The simplest way to obtain the approximate solutions $\tilde{y}_j^{(i)}(x)$ is to represent them in the form of power series. The unknown coefficients of the series could be obtained substituting these approximate solutions into equation (1) which is taken homogeneous to find functions $\tilde{y}_1^{(i)}(x)$ and $\tilde{y}_2^{(i)}(x)$ and non-homogeneous for the specific solution $\tilde{y}_R^{(i)}(x)$, and equating the coefficients before the same powers of $\tilde{x} = x - (x_{i+1} - x_i)/2$. This results in the following formulae for the functions $\tilde{y}_1^{(i)}(x)$, $\tilde{y}_2^{(i)}(x)$ and $\tilde{y}_R^{(i)}(x)$

$$\tilde{y}_1^{(i)}(x) = 1 - \frac{1}{2} G_0^{(i)} \tilde{x}^2 - \frac{1}{6} G_1^{(i)} \tilde{x}^3 + \dots \,, \tag{6a}$$

$$\tilde{y}_2^{(i)}(x) = \tilde{x} - \frac{1}{6} G_0^{(i)} \tilde{x}^3 + \dots \,, \tag{6b}$$

$$\tilde{y}_R^{(i)}(x) = \frac{1}{2} R_0^{(i)} \tilde{x}^2 + \frac{1}{6} R_1^{(i)} \tilde{x}^3 \dots \,. \tag{6c}$$

Here we have assumed the power expansions of the functions $G(x)$ and $R(x)$ in the form

$$G(x) = G_0^{(i)} + G_1^{(i)} \tilde{x} + \frac{1}{2} G_2^{(i)} \tilde{x}^2 + \dots \,, \tag{7a}$$

$$R(x) = R_0^{(i)} + R_1^{(i)} \tilde{x} + \frac{1}{2} R_2^{(i)} \tilde{x}^2 + \dots \,. \tag{7b}$$

Using such solutions is similar to employing the finite difference or finite element method of the corresponding order. For this reason this method with using polynomial functions has no evident advantages before the widely used standard methods.

In this paper we study another form of approximation of the solutions of the homogeneous equation:

$$\tilde{y}_{1,2}^{(i)}(x) = A_{1,2}^{(i)}(\tilde{x}) \exp\left(\Phi_{1,2}^{(i)}(\tilde{x})\right), \tag{8}$$

where $A_{1,2}^{(i)}(\tilde{x})$ and $\Phi_{1,2}^{(i)}(\tilde{x})$ are the polynomial functions of the following form:

$$A_{1,2}^{(i)}(\tilde{x}) = (A_0)_{1,2}^{(i)} + (A_1)_{1,2}^{(i)} \tilde{x} + \frac{1}{2}(A_2)_{1,2}^{(i)} \tilde{x}^2 + \ldots , \tag{9a}$$

$$\Phi_{1,2}^{(i)}(\tilde{x}) = (k_0)_{1,2}^{(i)} \tilde{x} + \frac{1}{2}(k_1)_{1,2}^{(i)} \tilde{x}^2 + \ldots . \tag{9b}$$

At the moment we will not pay attention to the form of the specific solution which should depend on the form of the function $R(x)$ in equation (1). For example, if the function $R(x)$ is sufficiently slowly varying at the segment $s_i$, a polynomial approximation (7b) is proper.

Now discuss the approximation (8), (9) in more detail. Such representation of the solutions makes it possible to describe rapidly oscillating or exponentially growing (decreasing) solutions at the segment $s_i$ even when the functions $A(\tilde{x})$ and $\Phi(\tilde{x})$ vary slowly at this segment. (Here and below we omit the indices of the solution number and the number of segment). Such situation takes place in the case when $G_0 h^2 \geq 1$, but $G_1 h^3 \ll 1$. In this situation the polynomial approximation of the solutions (6a), (6b) leads to large errors. When $G_0 h^2 \ll 1$ we expect that the polynomial-exponential solutions and the polynomial ones will behave similarly.

First it is necessary to show that such type of solutions can fulfill the homogeneous equation (1) with certain accuracy. After substituting them into the equation we obtain

$$A''(\tilde{x}) + 2A'(\tilde{x})k(\tilde{x}) + A(\tilde{x})k'(\tilde{x}) + \left[k^2(\tilde{x}) + G(\tilde{x})\right]A(\tilde{x}) = 0, \tag{10}$$

where prime denotes the derivative by $x$. Here we have introduced the function $k(\tilde{x}) = \Phi'(\tilde{x})$. The left-hand side of equation (10) is a polynomial and, therefore, to fulfill this equation with prescribed accuracy we have to nullify the coefficients before different powers of $\tilde{x}$ from the zero one up to the power corresponding to the degree of the approximation. This results in the following equations

$$\left(k_0^2 + G_0 + k_1\right)A_0 + 2k_0 A_1 + A_2 = 0 , \tag{11a}$$

$$\left(G_1 + 2k_0 k_1 + k_2\right)A_0 + \left(k_0^2 + G_0 + 3k_1\right)A_1 + 2k_0 A_2 + A_3 = 0 , \tag{11b}$$

... .

In practice we have to cut the series (9) at some maximum powers, $m_A$ for $A(\tilde{x})$ and $m_\Phi$ for $\Phi(\tilde{x})$, sufficient to obtain the prescribed degree of approximation. To determine the unknown coefficients $A_m$ and $k_m$ we require $m_A+m_\Phi$ equations. They can be picked up from the set (11). Note that since the set of equations (11) is homogeneous, one of the coefficients $A_m$, e.g. $A_0$, should be assumed to be known. Note that the set of equations (11) is linear in coefficients $A_m$ and non-linear in quantities $k_m$. In general case such system of equations is difficult to solve. We can simplify the problem decreasing the number of equations involved from (11), which reduces the degree of approximation. This creates some freedom in choosing the quantities $k_m$. The most simple way is to determine $k(\tilde{x})$ from the equation

$$k^2(\tilde{x}) = -G(\tilde{x}) , \tag{12}$$

which is similar to the zero-order WKB approximation for equation (1). In this case we use the first $m_A$ equations from (11) to find the coefficients $A_m$. For $m_A$=2 and $m_\Phi$=1 we can readily obtain the formulae

$$k_0 = \pm\sqrt{-G_0} , \quad k_1 = \pm\frac{G_1}{2\sqrt{-G_0}} , \tag{13}$$

$$A_0 = -(4G_0 + 3k_1)C , \quad A_1 = -2k_0 k_1 C , \quad A_2 = 3k_1^2 C , \tag{14}$$

where $C$ is an arbitrary constant.

Another method of obtaining the local solutions we illustrate for $m_A = m_\Phi = 1$. We shall use two first equations from the set (11). This leads to an additional condition for quantities $k_0$ and $k_1$:

$$3k_1^2 + 4k_1 G_0 + \left(k_0^2 + G_0\right)^2 - 2k_0 G_1 = 0 . \tag{15}$$

If we put $k_0^2 = -G_0$, which is the solution of equation (15) for $m_A=m_\Phi=0$ (zero-order approximation), then we obtain the quadratic equation which gives us $k_1$ (first-order approximation). Thus, in the framework of this scheme we have

$$k_0 = \pm\sqrt{-G_0} \ , \ k_1 = -\frac{2}{3}G_0\left[1-\sqrt{1\pm\frac{3}{2}\frac{G_1\sqrt{-G_0}}{G_0^2}}\right] , \qquad (16)$$

$$A_0 = 2k_0 C \ , \ A_1 = -k_1 C . \qquad (17)$$

From (16) follows that $(k_0)_1^{(i)} = -(k_0)_2^{(i)}$ but $(k_1)_1^{(i)} \neq -(k_1)_2^{(i)}$. This asymmetry of $k(\tilde{x})$ is in contrast with the WKB-type relation (12).

This scheme can be continued by consequent simultaneous increasing $m_A$ and $m_\Phi$ by unity. In this case equation (15) will be more complex. However, we will get a quadratic equation for the highest quantity $k_{m_A}$, if we substitute $k_m$ values from lower order approximations.

Another modification of the scheme described can be obtained using the requirement of symmetry of two roots of $k(\tilde{x})$, instead of the assumption $k_0^2 = -G_0$. In this case odd and even parts of equation (15) can be separated, which yields two equations for $k_0$ and $k_1$. Finally, we obtain the following formulae

$$k_0 = \pm\frac{4i}{\sqrt{3}}\frac{G_0^2}{G_1\left(1+\sqrt{1+\frac{16}{3}\frac{G_0^3}{G_1^2}}\right)} \ , \ k_1 = \pm\frac{2i}{\sqrt{3}}\frac{G_0}{1+\sqrt{1+\frac{16}{3}\frac{G_0^3}{G_1^2}}} , \qquad (18)$$

$$A_0 = 2k_0 C \ , \ A_1 = -\left(k_0^2 + G_0 + k_1\right)C . \qquad (19)$$

Further we shall call the local solution method in the form (13) and (14) as version 1, in the form (16) and (17) as version 2, and in the form (18) and (19) as version 3.

Note that all the local solutions of the form (8) considered above tend to the WKB solutions in the limit $|G_0^3| \gg |G_1^2|$. This makes it possible to use large-step mesh for numerical calculations even in the regions where WKB approximation is valid.

When using the local solutions in the above mentioned forms one should keep in mind that they either diverge or degenerate at the segment where $G_0=0$. Besides, there can arise some other points where the local solutions are degenerate. This is a disadvantage as compared with polynomial solutions which are always suitable. The points of degeneration can be found from analyzing the Wronskian for the solutions. For version 1 of the method the condition of Wronskian nullifying at the segment is

$$G_0\left(15G_1^2 + 64G_0^3\right) = 0 , \tag{20}$$

which yields again the condition $G_0=0$ and, besides this, a new degeneration point that appears at some negative $G_0$ value. Since here $\left|G_0^3\right| \sim \left|G_1^2\right|$, the point is situated in the region where the WKB approximation is not valid and the solution does not oscillate or vary rapidly. Thus at the segment of degeneration the polynomial solution can be used instead.

For version 2, the condition of Wronskian nullifying reads $G_0 = 0$. In this case no additional bad points arise.

For version 3, the condition is

$$G_0\left(3G_1^2 + 16G_0^3\right) = 0 . \tag{21}$$

Essentially, this condition is similar to the condition (19) except for numerical coefficients.

### 3. Numerical experiments

In this section, using numerical tests, we compare the versions of the local solution method proposed above with the standard finite difference method. For these tests we use the homogeneous Helmholtz equation (1) putting $R(x)=0$. In this case, non-trivial solutions of the equation appear owing to non-homogeneous boundary conditions. For the finite difference method [6] we employ the uniform mesh for which the finite difference scheme can be written in the following simple form

$$y_{i+1} - 2y_i + y_{i-1} + h^2 G(x_i) y_i = 0 , \tag{22}$$

where $h = x_i - x_{i-1}$.

The local solution method has been realized numerically following the procedure described in section 2. All three versions of the considered local solutions have been tested. The main part of calculations was performed for the function $G(x)=3x$. Thus, the solution can be expressed as a combination of Airy functions, $\text{Ai}(\sqrt[3]{3}x)$ and

Bi($\sqrt[3]{3}x$). Besides this, the function $G(x)=3x-0.06x^2$ has been used. The boundary conditions employed are the following

$$y|_{x=x_l} = 0, \quad y|_{x=x_r} = 1.\qquad(23)$$

As has been pointed out above, under certain conditions the local solutions in the exponential form can degenerate. We use two methods to overcome this difficulty. The first one is to use the polynomial local solutions at bad segments instead. The degeneration can be also avoided if we shift the end points of the bad segment. In the calculations we use both methods. In the case of version 1 the first method has been used. For version 2 we have used the second method. Remind that for this version the only case of degeneration is $G_0 \approx 0$. For version 3 we improve the segment with $G_0 \approx 0$ by the second method and use the first method at segments containing the additional points of degeneration.

As a quantitative characteristics of the calculation accuracy we introduce the local error

$$\delta y_i = y_i - y_{ex}(x_i),\qquad(24)$$

and the relative error

$$\delta = \sqrt{\frac{\sum_i (y_i - y_{ex}(x_i))^2}{\sum_i y_{ex}^2(x_i)}},\qquad(25)$$

where $y_{ex}(x)$ is the exact solution. As $y_{ex}(x)$ we have used a solution obtained at much finer uniform mesh. The summation in (25) is performed over the fine mesh nodes.

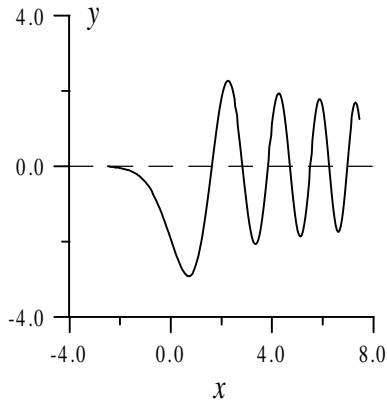

*Figure 2.* Solution of equation (1) for $G(x) = 3x$, $x_l = -2.5$, $x_r = 7.5$.

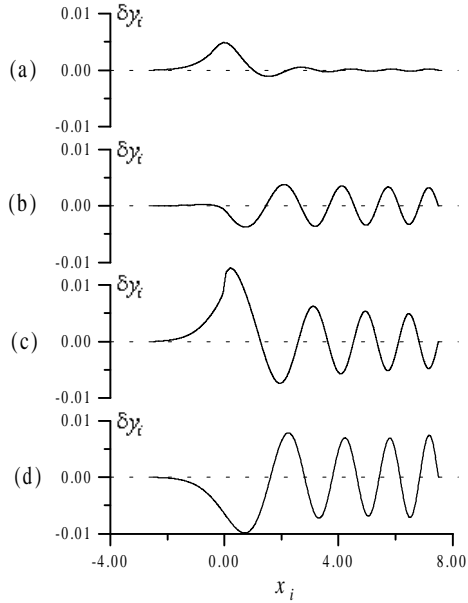

*Figure 3*. Error of numerical solution of equation (1) showed in figure 2, (a) - for version 3 of local solutions method, (b) - for version 2, (c) - for version 1 and (d) - for finite difference method.

Figures 4a, 4b and 4c display the relative errors as a function of the number of mesh nodes $N$ for all three versions of the local solution method. For comparison, the corresponding curve for the finite difference method is shown at every plot of figure 4. First of all, we have to note that all the versions demonstrate the convergence to the exact solution in average. The average rate of convergence is proportional to $1/N^2$ which corresponds to the degree of approximation and is similar to what we have in the case of the finite difference method. At the same time, all three versions yield more than two orders lower error level as compared to the finite difference method on the same mesh. A characteristic feature of the local solution method is non-monotonous convergence. The oscillations of the relative error are the largest in the case of version 1. For versions 2 and 3 these oscillations tend to vanish when $N$ increases.

We have also tested the method proposed for a solution having much more oscillations in the $x$ domain. In this case we have used another form of the $G$-function, the parabolic one: $G(x) = 3x - 0.06x^2$. For the calculations we choose $x_l = -2.5$, $x_r = 45.0$. The corresponding solution is shown in figure 5. It can be reproduced by the finite difference method. At $N=20000$ it yields the relative error $\delta = 2.1 \cdot 10^{-3}$. Version 3 of the local solution method yields nearly the same error $\delta = 2.2 \cdot 10^{-3}$ at $N=240$.

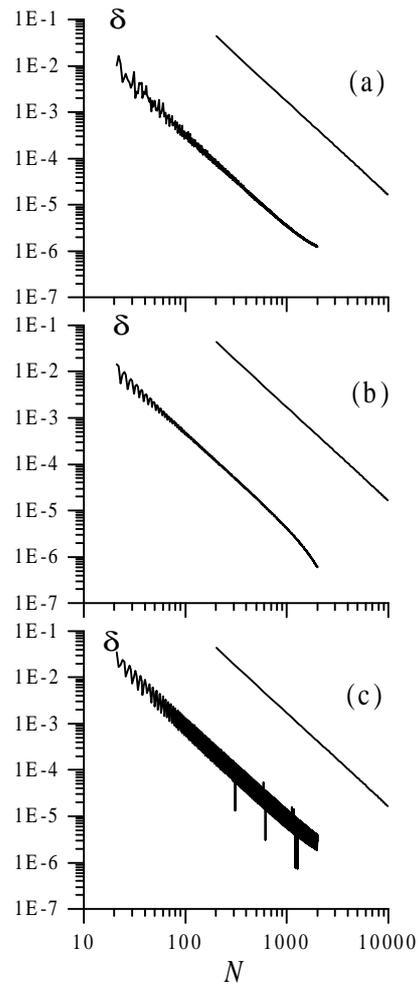

*Figure 4.* Relative error vs. number of mesh points, (a) - for version 3, (b) -for version 2 and (c) - for version1. In all three figures right curve displays relative error for finite difference method.

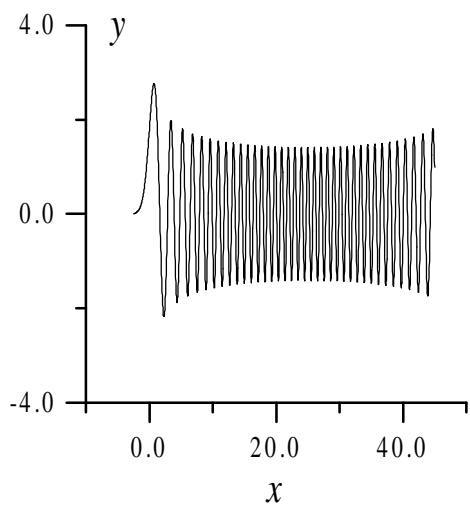

*Figure 5.* Solution of equation (1) for $G(x) = 3x - 0.06x^2$, $x_l = -2.5$, $x_r = 45$.

## 4. Conclusions.

In the presented paper we have analyzed and tested numerically the local solution method as applied to the Helmholtz equation. This method differs in principal from the standard methods of finite differences and finite elements. The method of local solutions makes it possible to use a wider class of basic functions approximating the solution, not only the polynomial ones. The most gain can be obtained when the basic functions are close to exact solutions. In the present paper we have considered the basic functions of the polynomial-exponential type. The form of these functions is close to the WKB solutions. Owing to this we have obtained good numerical accuracy if the solutions are rapidly oscillating or exponentially increasing (decreasing). Besides, these basic functions are also able to reproduce the solutions with prescribed accuracy in regions of slow solution variation. This allows one to apply the method of local solutions to general problems.

The numerical tests performed have demonstrated convergence of the method proposed. In average the convergence corresponds to the degree of approximation, although it is not monotonous general. The method proposed has shown the evident advantage before the standard finite difference method of the same order for modelling solutions of the Airy equation and the parabolic-cylinder-type equation. Depending on the solution character, the local solution method requires 10-100 times rarer mesh than finite difference method to obtain the same accuracy.